\documentclass[11pt,a4paper]{article}
\usepackage{jhepstyle}
\usepackage[utf8]{inputenc}
\usepackage{amsmath}
\usepackage{amsthm}
\usepackage{amssymb}
\usepackage{enumitem}   
\usepackage[english]{babel}
\usepackage{url}
\usepackage{mathtools}
\usepackage{bbold}
\usepackage{slashed}
\usepackage{comment}

\usepackage{subcaption}
\usepackage{tikz}
\usetikzlibrary{arrows,shapes.misc,decorations.markings,decorations.pathmorphing,positioning,intersections}
\tikzset{cross/.style={cross out, draw=black, minimum size=2*(#1-\pgflinewidth), inner sep=0pt, outer sep=0pt},
cross/.default={1.5mm}}
\tikzset{mydash/.style={dashed, dash pattern=on 4pt off 5pt}}
\tikzset{
  vertex/.style={draw,shape=circle,fill=black,minimum size=3pt,inner sep=0pt},
  cross/.style={cross out, draw=black,thick, minimum size=6pt, inner sep=0pt, outer sep=0pt},
  external/.style={inner sep=2pt},
  plabel/.style={inner sep=2pt},
  blob/.style={circle,fill=black!20,minimum size=0.7cm,draw,thick},
  whiteblob/.style={circle,fill=white,minimum size=1.0cm,draw,thick},
  effective/.style={rectangle,fill=black!20,minimum size=0.5cm,draw,thick},
  vev/.style={shape=vev,draw,inner sep=2pt,thick},
  mass/.style={shape=cross,draw,thick},
  rscalar/.style={dashed,thick},
  mfermion/.style={thick},
  scalar/.style={postaction={decorate}, decoration={markings,mark=at position .55 with {\arrow{latex}}},dashed,thick},
  ooscalar/.style={postaction={decorate}, decoration={markings,mark=at position .7 with {\arrow{latex}}},dashed,thick},
  fermion/.style={postaction={decorate}, decoration={markings,mark=at position .55 with {\arrow{latex}}},thick},
  majfermion/.style={postaction={decorate}, decoration={markings,mark=at position .7 with {\arrow{latex}}},thick},
  oofermion/.style={postaction={decorate}, decoration={markings,mark=at position .85 with {\arrow{latex}}, mark=at position .35 with {\arrowreversed{latex}}},thick},
  iifermion/.style={postaction={decorate}, decoration={markings,mark=at position .35 with {\arrowreversed{latex}}, mark=at position .85 with {\arrow{latex}}},thick},
  gaugeboson/.style={decorate, decoration={snake},thick},
  gluon/.style={decorate, decoration={coil,amplitude=4pt, segment length=5pt},thick},
  photon/.style={decorate, decoration={snake},thick},
  dashdot/.style={dash pattern=on .4pt off 3pt on 4pt off 3pt,thick}
}

\begin{document}

\title{The abelian gauge-Yukawa $\beta$-functions at large $N_f$}

\author{Tommi Alanne,}
\author{Simone Blasi}

\affiliation{Max-Planck-Institut f\"{u}r Kernphysik, Saupfercheckweg 1, 69117 Heidelberg, Germany}

\emailAdd{tommi.alanne@mpi-hd.mpg.de}
\emailAdd{simone.blasi@mpi-hd.mpg.de}

\abstract{
We study the impact of the Yukawa interaction in the large-$N_f$ limit to
the abelian gauge theory. We compute the coupled $\beta$-functions for the system
in a closed form at $\mathcal{O}(1/N_f)$.
}

\maketitle
\newpage

\section{Introduction}
A comprehensive understanding of the UV behaviour of gauge-Yukawa theories has become of
key importance with the growing interest in the asymptotic-safety 
paradigm~\cite{Litim:2014uca,Mann:2017wzh,Pelaggi:2017abg,Antipin:2017ebo}. Prime candidates 
for these considerations are gauge-Yukawa models with a large number of fermion flavours, $N_f$. Computing the
leading large-$N_f$ contribution to the $\beta$-functions was pioneered by evaluating the 
$\mathcal{O}(1/N_f)$ gauge $\beta$-functions~\cite{Espriu:1982pb,PalanquesMestre:1983zy,Gracey:1996he} 
for $N_f$ fermions charged under the gauge group; see also Refs~\cite{Holdom:2010qs,Shrock:2013cca}. 

We recently computed the $\mathcal{O}(1/N_f)$ $\beta$-function 
for Yukawa-theory~\cite{Alanne:2018ene}
inspired by the earlier works~\cite{Kowalska:2017pkt,Antipin:2018zdg}. 
The Yukawa-theory is closely related to the
Gross--Neveu model, which has been extensively studied in the past using a different approach; see e.g. 
Refs~\cite{Gracey:1990wi,Vasiliev:1992wr,Gracey:1993kb,Gracey:1993kc}. For Gross--Neveu--Yukawa model the behaviour near the fixed point in terms of 
critical exponents is known
up to $\mathcal{O}(1/N_f^2)$~\cite{Gracey:2017fzu,Manashov:2017rrx}.  However, the strength of our analysis is that 
we readily achieved a closed form expression of the $\beta$-function, and as shown in the present work,
the procedure is straighforwardly generalisable to include gauge interactions. 

In this paper, we compute the leading $1/N_f$ contributions to the $\beta$-functions of the gauge-Yukawa system 
in a closed form. This result is new and sheds light to the impact of the Yukawa interaction to the gauge theory
in the large-$N_f$ limit. 

The gauge contribution to the Yukawa $\beta$-function was computed in the abelian case in Ref.~\cite{Kowalska:2017pkt} and later 
generalised to non-abelian and semi-simple gauge groups in Ref.~\cite{Antipin:2018zdg}
assuming that only one flavour of fermions couples
to the scalar via Yukawa interaction. We relax this assumption and show that it is possible to get a closed form expressions also in the general case. 
The current result provides a groundwork for several interesting extensions including e.g.
non-abelian gauge groups and chiral fermions.

The paper is organized as follows: In Sec.~\ref{sec:def} we introduce the framework and notations and in 
Sec.~\ref{sec:renC} compute the new contributions to the renormalization constants and $\beta$-functions. In Sec.~\ref{sec:beta} we collect the 
results and comment on the structure of the coupled system, and 
in Sec.~\ref{sec:concl} we conclude.

\section{The framework}
\label{sec:def}

We consider the massless U(1) gauge theory with $N_f$ fermion flavours (QED) with a gauge-singlet 
real scalar field coupling to the fermionic multiplet, $\psi$, via 
Yukawa interaction:
\begin{equation}\label{eq:Lyuk}
 \mathcal{L} = -\frac{1}{4}F_{\mu\nu}F^{\mu\nu}+\frac{1}{2}\partial_{\mu}\phi\partial^{\mu}\phi
    +i \bar{\psi}\slashed{D}\psi + y  \bar{\psi} \psi \phi.
\end{equation}
We define the rescaled gauge and Yukawa couplings,  
\begin{equation}
 E \equiv \frac{e^2}{4 \pi^2} N_f,\ \text{and}\
 K \equiv \frac{y^2}{4 \pi^2} N_f,
\end{equation}
which are kept constant in the limit $N_f\to\infty$.
The purpose of this paper is to derive  the coupled
system of $\beta$-functions for $E$ and $K$ at the $1/N_f$ level:
\begin{align}
    \label{eq:coupledE}
    \beta_E &\equiv \frac{\mathrm{d}E}{\mathrm{d}\ln\mu}
	= E \left( K \frac{\partial}{\partial K} + E \frac{\partial}{\partial E} \right) G_1(K,E),\\
    \label{eq:coupledK}
    \beta_K &\equiv \frac{\mathrm{d}K}{\mathrm{d}\ln\mu}
	= K \left( K \frac{\partial}{\partial K} + E \frac{\partial}{\partial E} \right) H_1(K,E),
\end{align}
where $G_1$ and $H_1$ are defined by 
\begin{align}\label{eq:ZK}
\log Z_E&\equiv \log Z_3^{-1} = \sum_{n=1}^\infty \frac{G_n(K,E)}{\epsilon^n},\\
\log Z_K &\equiv \log (Z_S^{-1} Z_F^{-2} Z_V^2) = \sum_{n=1}^\infty \frac{H_n(K,E)}{\epsilon^n},
\end{align}
and $Z_3$, $Z_S$, $Z_F$, and $Z_V$ are the renormalization constants for the photon, the scalar,
and the fermion wave function, and the 1PI vertex, respectively.
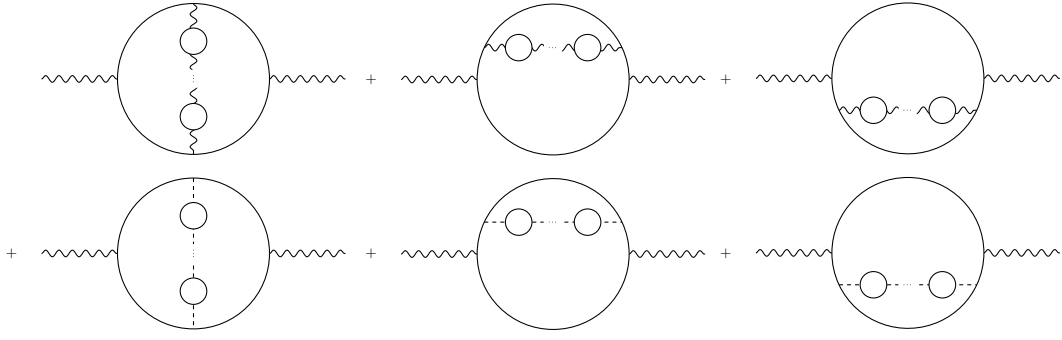
\begin{figure}[t]
    \scalebox{0.5}{
\hspace{0.8cm}\hphantom{$+$} \hspace{.5cm}\vspace{2cm}
\begin{minipage}[c]{.25\textwidth}
  \begin{tikzpicture}[node distance=2cm]
    \coordinate (v1);
    \coordinate[right = of v1] (v2);
    \coordinate[right = of v2] (v3);
    \coordinate[right = of v3] (v4);
    \coordinate[right = of v4] (v5);
    \draw[photon] (v1) -- (v2);
    \draw[mfermion] (v2) arc(180:90:2) coordinate (v8) arc(90:0:2);
    \draw[mfermion] (v2) arc(-180:-90:2) coordinate (v9) arc(-90:0:2);
    \draw[photon] (v4) -- (v5);
    \draw[photon] (v8) --  (v9) 
	node[pos=0.25,draw,solid,whiteblob,minimum size=0.7 cm] {}
	node[pos=0.44] (v11) {}
	node[pos=0.56] (v12) {}
	node[midway,circle,fill=white,dotted, minimum size=0.5 cm]{}
	node[pos=0.75,draw,solid,whiteblob,minimum size=0.7 cm] {};
    \draw[dotted] (v11)--(v12);
  \end{tikzpicture}
\end{minipage}
  ~\hspace{4.5cm}$+$\hspace{0.5cm}
\begin{minipage}[c]{.25\textwidth}
    \vspace{-.25cm}
  \begin{tikzpicture}[node distance=2cm]
    \coordinate (v1);
    \coordinate[right = of v1] (v2);
    \coordinate[right = of v2] (v3);
    \coordinate[right = of v3] (v4);
    \coordinate[right = of v4] (v5);
    \draw[photon] (v1) -- (v2);
    \draw[mfermion] (v2) arc(180:155:2) coordinate (v8) arc(155:25:2) coordinate (v9) arc(25:0:2);
    \draw[mfermion] (v2) arc(-180:0:2);
    \draw[photon] (v4) -- (v5);
    \draw[photon] (v8) --  (v9) 
	node[pos=0.25,draw,solid,whiteblob,minimum size=0.7 cm] {}
	node[pos=0.43] (v11) {}
	node[pos=0.57] (v12) {}
	node[midway,circle,fill=white,dotted, minimum size=0.5 cm]{}
	node[pos=0.75,draw,solid,whiteblob,minimum size=0.7 cm] {};
    \draw[dotted] (v11)--(v12);
  \end{tikzpicture}
\end{minipage}
  \hspace{4.5cm}$+$ \hspace{.5cm}\vspace{2cm}
\begin{minipage}[c]{.25\textwidth}
\vspace{0.25cm}
  \begin{tikzpicture}[node distance=2cm]
    \coordinate (v1);
    \coordinate[right = of v1] (v2);
    \coordinate[right = of v2] (v3);
    \coordinate[right = of v3] (v4);
    \coordinate[right = of v4] (v5);
    \draw[photon] (v1) -- (v2);
    \draw[mfermion] (v2) arc(-180:-155:2) coordinate (v8) arc(-155:-25:2) coordinate (v9) arc(-25:0:2);
    \draw[mfermion] (v2) arc(180:0:2);
    \draw[photon] (v4) -- (v5);
    \draw[photon] (v8) --  (v9) 
	node[pos=0.25,draw,solid,whiteblob,minimum size=0.7 cm] {}
	node[pos=0.43] (v11) {}
	node[pos=0.57] (v12) {}
	node[midway,circle,fill=white,dotted, minimum size=0.5 cm]{}
	node[pos=0.75,draw,solid,whiteblob,minimum size=0.7 cm] {};
    \draw[dotted] (v11)--(v12);
  \end{tikzpicture}
\end{minipage}
}\\
    \scalebox{0.5}{
\hspace{0.8cm}$+$ \hspace{.5cm}\vspace{2cm}
  \begin{minipage}[c]{.25\textwidth}
  \begin{tikzpicture}[node distance=2cm]
    \coordinate (v1);
    \coordinate[right = of v1] (v2);
    \coordinate[right = of v2] (v3);
    \coordinate[right = of v3] (v4);
    \coordinate[right = of v4] (v5);
    \draw[photon] (v1) -- (v2);
    \draw[mfermion] (v2) arc(180:90:2) coordinate (v8) arc(90:0:2);
    \draw[mfermion] (v2) arc(-180:-90:2) coordinate (v9) arc(-90:0:2);
    \draw[photon] (v4) -- (v5);
    \draw[rscalar] (v8) --  (v9) 
	node[pos=0.25,draw,solid,whiteblob,minimum size=0.7 cm] {}
	node[pos=0.44] (v11) {}
	node[pos=0.56] (v12) {}
	node[midway,circle,fill=white,dotted, minimum size=0.5 cm]{}
	node[pos=0.75,draw,solid,whiteblob,minimum size=0.7 cm] {};
    \draw[dotted] (v11)--(v12);
  \end{tikzpicture}
\end{minipage}
  ~\hspace{4.5cm}$+$\hspace{0.5cm}
\begin{minipage}[c]{.25\textwidth}
    \vspace{-.25cm}
  \begin{tikzpicture}[node distance=2cm]
    \coordinate (v1);
    \coordinate[right = of v1] (v2);
    \coordinate[right = of v2] (v3);
    \coordinate[right = of v3] (v4);
    \coordinate[right = of v4] (v5);
    \draw[photon] (v1) -- (v2);
    \draw[mfermion] (v2) arc(180:155:2) coordinate (v8) arc(155:25:2) coordinate (v9) arc(25:0:2);
    \draw[mfermion] (v2) arc(-180:0:2);
    \draw[photon] (v4) -- (v5);
    \draw[rscalar] (v8) --  (v9) 
	node[pos=0.25,draw,solid,whiteblob,minimum size=0.7 cm] {}
	node[pos=0.43] (v11) {}
	node[pos=0.57] (v12) {}
	node[midway,circle,fill=white,dotted, minimum size=0.5 cm]{}
	node[pos=0.75,draw,solid,whiteblob,minimum size=0.7 cm] {};
    \draw[dotted] (v11)--(v12);
  \end{tikzpicture}
\end{minipage}
  \hspace{4.5cm}$+$ \hspace{.5cm}\vspace{2cm}
\begin{minipage}[c]{.25\textwidth}
\vspace{0.25cm}
  \begin{tikzpicture}[node distance=2cm]
    \coordinate (v1);
    \coordinate[right = of v1] (v2);
    \coordinate[right = of v2] (v3);
    \coordinate[right = of v3] (v4);
    \coordinate[right = of v4] (v5);
    \draw[photon] (v1) -- (v2);
    \draw[mfermion] (v2) arc(-180:-155:2) coordinate (v8) arc(-155:-25:2) coordinate (v9) arc(-25:0:2);
    \draw[mfermion] (v2) arc(180:0:2);
    \draw[photon] (v4) -- (v5);
    \draw[rscalar] (v8) --  (v9) 
	node[pos=0.25,draw,solid,whiteblob,minimum size=0.7 cm] {}
	node[pos=0.43] (v11) {}
	node[pos=0.57] (v12) {}
	node[midway,circle,fill=white,dotted, minimum size=0.5 cm]{}
	node[pos=0.75,draw,solid,whiteblob,minimum size=0.7 cm] {};
    \draw[dotted] (v11)--(v12);
  \end{tikzpicture}
\end{minipage}
}
  \caption{Photon self-energy corrections.}
  \label{fig:photonSE}
\end{figure}


The photon wave function renormalization constant, $Z_3$, is given by
\begin{equation}\label{eq:Z_3def}
 Z_3 = 1 - \text{div}
 \left\{Z_3\Pi_0(p^2, Z_K K, Z_E E, \epsilon) \right\},
\end{equation}
where $\Pi_0$ is the self-energy divided by the external momentum squared, $p^2$,  
and we denote the poles of $X$ in $\epsilon$ by $\text{div}{X}$. 
The self-energy can be written as
\begin{equation}\label{eq:pi0}
\begin{split}
 &\Pi_0(p^2,K_0,E_0,\epsilon)\\
 &\qquad= E_0 \, \Pi_E^{(1)}(p^2,\epsilon)
 + \frac{1}{N_f} \sum_{n=2}^\infty \left( E_0^n \Pi_E^{(n)}(p^2,\epsilon)
 + E_0 K_0^{n-1} \Pi_K^{(n)}(p^2,\epsilon)\right)+\mathcal{O}(1/N_f^2),
 \end{split}
\end{equation}
where $\Pi_E^{(1)}$ is the one-loop contribution,
and $\Pi_E^{(n)}$ and $\Pi_K^{(n)}$ contain the $n$-loop part 
consisting of $n-2$ fermion bubbles 
in the gauge and Yukawa chains summing over the topologies given in Fig.~\ref{fig:photonSE}. 

\begin{figure}[t]
    \scalebox{0.5}{
\hspace{0.8cm}\hphantom{$+$} \hspace{.5cm}\vspace{2cm}
\begin{minipage}[c]{.25\textwidth}
  \begin{tikzpicture}[node distance=2cm]
    \coordinate (v1);
    \coordinate[right = of v1] (v2);
    \coordinate[right = of v2] (v3);
    \coordinate[right = of v3] (v4);
    \coordinate[right = of v4] (v5);
    \draw[rscalar] (v1) -- (v2);
    \draw[mfermion] (v2) arc(180:90:2) coordinate (v8) arc(90:0:2);
    \draw[mfermion] (v2) arc(-180:-90:2) coordinate (v9) arc(-90:0:2);
    \draw[rscalar] (v4) -- (v5);
    \draw[rscalar] (v8) --  (v9) 
	node[pos=0.12,draw,solid,whiteblob,minimum size=0.5 cm] {}
	node[pos=0.32,draw,solid,whiteblob,minimum size=0.5 cm] {}
	node[pos=0.44] (v11) {}
	node[pos=0.56] (v12) {}
	node[midway,circle,fill=white,dotted, minimum size=0.5 cm]{}
	node[pos=0.68,draw,solid,whiteblob,minimum size=0.5 cm] {}
	node[pos=0.88,draw,solid,whiteblob,minimum size=0.5 cm] {};
    \draw[dotted] (v11)--(v12);
  \end{tikzpicture}
\end{minipage}
  ~\hspace{4.5cm}$+$\hspace{0.5cm}
\begin{minipage}[c]{.25\textwidth}
    \vspace{-.25cm}
  \begin{tikzpicture}[node distance=2cm]
    \coordinate (v1);
    \coordinate[right = of v1] (v2);
    \coordinate[right = of v2] (v3);
    \coordinate[right = of v3] (v4);
    \coordinate[right = of v4] (v5);
    \draw[rscalar] (v1) -- (v2);
    \draw[mfermion] (v2) arc(180:155:2) coordinate (v8) arc(155:25:2) coordinate (v9) arc(25:0:2);
    \draw[mfermion] (v2) arc(-180:0:2);
    \draw[rscalar] (v4) -- (v5);
    \draw[rscalar] (v8) --  (v9) 
	node[pos=0.12,draw,solid,whiteblob,minimum size=0.5 cm] {}
	node[pos=0.32,draw,solid,whiteblob,minimum size=0.5 cm] {}
	node[pos=0.43] (v11) {}
	node[pos=0.57] (v12) {}
	node[midway,circle,fill=white,dotted, minimum size=0.5 cm]{}
	node[pos=0.68,draw,solid,whiteblob,minimum size=0.5 cm] {}
	node[pos=0.88,draw,solid,whiteblob,minimum size=0.5 cm] {};
    \draw[dotted] (v11)--(v12);
  \end{tikzpicture}
\end{minipage}
  \hspace{4.5cm}$+$ \hspace{.5cm}\vspace{2cm}
\begin{minipage}[c]{.25\textwidth}
\vspace{0.25cm}
  \begin{tikzpicture}[node distance=2cm]
    \coordinate (v1);
    \coordinate[right = of v1] (v2);
    \coordinate[right = of v2] (v3);
    \coordinate[right = of v3] (v4);
    \coordinate[right = of v4] (v5);
    \draw[rscalar] (v1) -- (v2);
    \draw[mfermion] (v2) arc(-180:-155:2) coordinate (v8) arc(-155:-25:2) coordinate (v9) arc(-25:0:2);
    \draw[mfermion] (v2) arc(180:0:2);
    \draw[rscalar] (v4) -- (v5);
    \draw[rscalar] (v8) --  (v9) 
	node[pos=0.12,draw,solid,whiteblob,minimum size=0.5 cm] {}
	node[pos=0.32,draw,solid,whiteblob,minimum size=0.5 cm] {}
	node[pos=0.43] (v11) {}
	node[pos=0.57] (v12) {}
	node[midway,circle,fill=white,dotted, minimum size=0.5 cm]{}
	node[pos=0.68,draw,solid,whiteblob,minimum size=0.5 cm] {}
	node[pos=0.88,draw,solid,whiteblob,minimum size=0.5 cm] {};
    \draw[dotted] (v11)--(v12);
  \end{tikzpicture}
\end{minipage}
}\\
    \scalebox{0.5}{
 \hspace{0.7cm} $+$ \hspace{.5cm}\vspace{2cm}
\begin{minipage}[c]{.25\textwidth}
  \begin{tikzpicture}[node distance=2cm]
    \coordinate (v1);
    \coordinate[right = of v1] (v2);
    \coordinate[right = of v2] (v3);
    \coordinate[right = of v3] (v4);
    \coordinate[right = of v4] (v5);
    \draw[rscalar] (v1) -- (v2);
    \draw[mfermion] (v2) arc(180:90:2) coordinate (v8) arc(90:0:2);
    \draw[mfermion] (v2) arc(-180:-90:2) coordinate (v9) arc(-90:0:2);
    \draw[rscalar] (v4) -- (v5);
    \draw[photon] (v8) --  (v9) 
	node[pos=0.25,draw,solid,whiteblob,minimum size=0.7 cm] {}
	node[pos=0.44] (v11) {}
	node[pos=0.56] (v12) {}
	node[midway,circle,fill=white,dotted, minimum size=0.5 cm]{}
	node[pos=0.75,draw,solid,whiteblob,minimum size=0.7 cm] {};
    \draw[dotted] (v11)--(v12);
  \end{tikzpicture}
\end{minipage}
  ~\hspace{4.5cm}$+$\hspace{0.5cm}
\begin{minipage}[c]{.25\textwidth}
    \vspace{-.25cm}
  \begin{tikzpicture}[node distance=2cm]
    \coordinate (v1);
    \coordinate[right = of v1] (v2);
    \coordinate[right = of v2] (v3);
    \coordinate[right = of v3] (v4);
    \coordinate[right = of v4] (v5);
    \draw[rscalar] (v1) -- (v2);
    \draw[mfermion] (v2) arc(180:155:2) coordinate (v8) arc(155:25:2) coordinate (v9) arc(25:0:2);
    \draw[mfermion] (v2) arc(-180:0:2);
    \draw[rscalar] (v4) -- (v5);
    \draw[photon] (v8) --  (v9) 
	node[pos=0.25,draw,solid,whiteblob,minimum size=0.7 cm] {}
	node[pos=0.43] (v11) {}
	node[pos=0.57] (v12) {}
	node[midway,circle,fill=white,dotted, minimum size=0.5 cm]{}
	node[pos=0.75,draw,solid,whiteblob,minimum size=0.7 cm] {};
    \draw[dotted] (v11)--(v12);
  \end{tikzpicture}
\end{minipage}
  \hspace{4.5cm}$+$ \hspace{.5cm}\vspace{2cm}
\begin{minipage}[c]{.25\textwidth}
\vspace{0.25cm}
  \begin{tikzpicture}[node distance=2cm]
    \coordinate (v1);
    \coordinate[right = of v1] (v2);
    \coordinate[right = of v2] (v3);
    \coordinate[right = of v3] (v4);
    \coordinate[right = of v4] (v5);
    \draw[rscalar] (v1) -- (v2);
    \draw[mfermion] (v2) arc(-180:-155:2) coordinate (v8) arc(-155:-25:2) coordinate (v9) arc(-25:0:2);
    \draw[mfermion] (v2) arc(180:0:2);
    \draw[rscalar] (v4) -- (v5);
    \draw[photon] (v8) --  (v9) 
	node[pos=0.25,draw,solid,whiteblob,minimum size=0.7 cm] {}
	node[pos=0.43] (v11) {}
	node[pos=0.57] (v12) {}
	node[midway,circle,fill=white,dotted, minimum size=0.5 cm]{}
	node[pos=0.75,draw,solid,whiteblob,minimum size=0.7 cm] {};
    \draw[dotted] (v11)--(v12);
  \end{tikzpicture}
\end{minipage}
  }
  \caption{Scalar self-energy corrections.}
  \label{fig:scalarSE}
\end{figure}
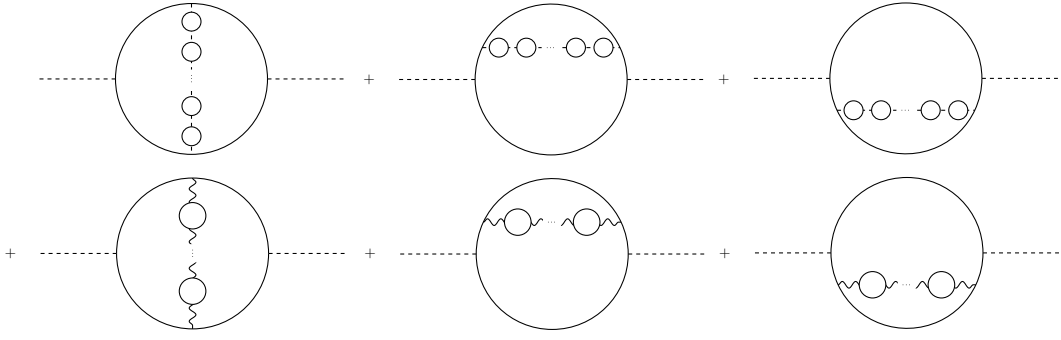

The scalar wave function renormalization constant, $Z_S$, is determined via
\begin{equation}
    \label{eq:Z_Sdef}
    Z_S = 1 - \text{div}\left\{Z_S S_0(p^2,Z_K K, Z_E E,\epsilon) \right\}, 
\end{equation}
with the scalar self-energy given by
\begin{equation}
\begin{split}
    &S_0(p^2,K_0,E_0,\epsilon)\\
    &\qquad= K_0 S_K^{(1)}(p^2,\epsilon) 
    +\frac{1}{N_f} \sum_{n=2}^\infty \left( K_0^n S_K^{(n)}(p^2,\epsilon)
    + K_0 E_0^{n-1} S_E^{(n)}(p^2,\epsilon) \right)+\mathcal{O}(1/N_f^2),
\end{split}
\end{equation}
where $S_K^{(1)}$ is the one-loop result, 
and $S_{K}^{(n)}$ and $S_{E}^{(n)}$ the $n$-loop terms consisting of
$n-2$ fermion bubbles in the Yukawa and gauge 
chains summing over the topologies shown in Fig.~\ref{fig:scalarSE}.

\begin{figure}[t]
\centering
\begin{subfigure}{0.4\textwidth}\quad
~\hspace{-1.55cm}
    \scalebox{0.75}{
\begin{minipage}[c]{.25\textwidth}
\vspace{0.25cm}
  \begin{tikzpicture}[node distance=1cm]
    \coordinate (v1);
    \coordinate[right = of v1] (v2);
    \coordinate[right = of v2] (v31);
    \coordinate[right = of v31] (v32);
    \coordinate[right = of v32] (v3);
    \coordinate[right = of v3] (v4);
    \draw[mfermion] (v1) -- (v4);
    \draw[rscalar] (v2) arc(180:0:1.5) 
	node[pos=0.12,draw,solid,whiteblob,minimum size=0.5 cm] {}
	node[pos=0.32,draw,solid,whiteblob,minimum size=0.5 cm] {}
	node[pos=0.45] (v11) {}
	node[pos=0.55] (v12) {}
	node[midway,circle,fill=white,dotted, minimum size=0.5 cm]{}
	node[pos=0.68,draw,solid,whiteblob,minimum size=0.5 cm] {}
	node[pos=0.88,draw,solid,whiteblob,minimum size=0.5 cm] {};
    \draw[dotted] (v11)--(v12);
  \end{tikzpicture}
\end{minipage}
  \hspace{3.8cm}$+$ \hspace{.5cm}\vspace{2cm}
\begin{minipage}[c]{.25\textwidth}
\vspace{0.25cm}
  \begin{tikzpicture}[node distance=1cm]
    \coordinate (v1);
    \coordinate[right = of v1] (v2);
    \coordinate[right = of v2] (v31);
    \coordinate[right = of v31] (v32);
    \coordinate[right = of v32] (v3);
    \coordinate[right = of v3] (v4);
    \draw[mfermion] (v1) -- (v4);
    \draw[photon] (v2) arc(180:0:1.5) 
	node[pos=0.25,draw,solid,whiteblob,minimum size=0.6 cm] {}
	node[pos=0.45] (v11) {}
	node[pos=0.55] (v12) {}
	node[midway,circle,fill=white,dotted, minimum size=0.5 cm]{}
	node[pos=0.75,draw,solid,whiteblob,minimum size=0.6 cm] {};
    \draw[dotted] (v11)--(v12);
  \end{tikzpicture}
\end{minipage}
  }
  \caption{Fermion self-energy corrections.}
  \label{fig:fermionSE}
\end{subfigure}\\
~\\
~\\
~\\
\begin{subfigure}{0.4\textwidth}
\vspace{0.25cm}
~\hspace{-1.25cm}
    \scalebox{0.75}{
\begin{minipage}[c]{.25\textwidth}
  \begin{tikzpicture}[node distance=1cm]
    \coordinate (v1);
    \coordinate[right = of v1] (v21);
    \coordinate[right = of v21] (v2);
    \coordinate[above right = of v2] (v31);
    \coordinate[above right = of v31] (v32);
    \coordinate[above right = of v32] (v3);
    \coordinate[below right = of v2] (v41);
    \coordinate[below right = of v41] (v42);
    \coordinate[below right = of v42] (v4);
    \draw[rscalar] (v1) -- (v2);
    \draw[mfermion] (v2) -- (v3) node[pos=0.85] (v5) {};
    \draw[mfermion] (v2) -- (v4) node[pos=0.85] (v6) {};
    \draw[rscalar] (v5) --  (v6) 
	node[pos=0.12,draw,solid,whiteblob,minimum size=0.5 cm] {}
	node[pos=0.32,draw,solid,whiteblob,minimum size=0.5 cm] {}
	node[pos=0.43] (v11) {}
	node[pos=0.57] (v12) {}
	node[midway,circle,fill=white,dotted, minimum size=0.5 cm]{}
	node[pos=0.68,draw,solid,whiteblob,minimum size=0.5 cm] {}
	node[pos=0.88,draw,solid,whiteblob,minimum size=0.5 cm] {};
    \draw[dotted] (v11)--(v12);
  \end{tikzpicture}
\end{minipage}
  \hspace{3.8cm}$+$ \hspace{.5cm}\vspace{2cm}
\begin{minipage}[c]{.25\textwidth}
\vspace{0.25cm}
  \begin{tikzpicture}[node distance=1cm]
    \coordinate (v1);
    \coordinate[right = of v1] (v21);
    \coordinate[right = of v21] (v2);
    \coordinate[above right = of v2] (v31);
    \coordinate[above right = of v31] (v32);
    \coordinate[above right = of v32] (v3);
    \coordinate[below right = of v2] (v41);
    \coordinate[below right = of v41] (v42);
    \coordinate[below right = of v42] (v4);
    \draw[rscalar] (v1) -- (v2);
    \draw[mfermion] (v2) -- (v3) node[pos=0.85] (v5) {};
    \draw[mfermion] (v2) -- (v4) node[pos=0.85] (v6) {};
    \draw[photon] (v5) --  (v6) 
	node[pos=0.22,draw,solid,whiteblob,minimum size=0.6 cm] {}
	node[pos=0.43] (v11) {}
	node[pos=0.57] (v12) {}
	node[midway,circle,fill=white,dotted, minimum size=0.5 cm]{}
	node[pos=0.78,draw,solid,whiteblob,minimum size=0.6 cm] {};
    \draw[dotted] (v11)--(v12);
  \end{tikzpicture}
\end{minipage}
  }
  \caption{Vertex corrections.}
  \label{fig:vertex}
\end{subfigure}
\caption{Gauge and Yukawa contributions to fermion self-energy and the  vertex corrections due 
to a chain of fermion bubbles.}
\label{fig:bubbleChains}
\end{figure}
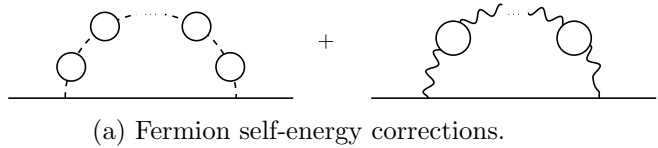
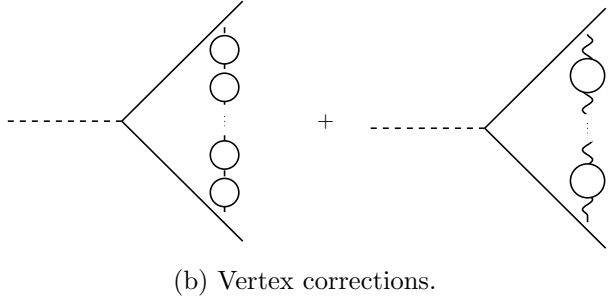

For the fermion self-energy and vertex renormalization constants, the lowest non-trivial 
contributions are already $\mathcal{O}(1/N_f)$, and we have
\begin{align}\label{eq:Z_fdef}
    Z_f =& 1 - \text{div}\left\{\Sigma_0(p^2,Z_K K, Z_E E, \epsilon) \right\},\\
    \Sigma_0(p^2,K_0,E_0,\epsilon)
    =& 1 +\frac{1}{N_f} 
    \sum_{n=1}^\infty \left( K_0^n \Sigma_K^{(n)}(p^2,\epsilon)
  + E_0^n \Sigma^{(n)}_E(p^2,\epsilon) \right)+\mathcal{O}(1/N_f^2),
\end{align}
where $\Sigma^{(n)}_{K}$ and $\Sigma^{(n)}_{E}$ 
are depicted in Fig.~\ref{fig:fermionSE} with $n-1$ fermion bubbles. 
Similarly,
\begin{align}\label{eq:Z_Vdef}
    Z_V &= 1 - \text{div}\left\{V_0(p^2,Z_K K, Z_E E, \epsilon) \right\},\\
    V_0(p^2,K_0,E_0,\epsilon)
    &= 1 + \frac{1}{N_f}\sum_{n=1}^\infty \left( K_0^n V_K^{(n)}(p^2,\epsilon)
    + E_0^n V_E^{(n)}(p^2,\epsilon) \right)+\mathcal{O}(1/N_f^2),
\end{align}
where $V_{K}^{(n)}$ and  $V_{E}^{(n)}$ contain
$n-1$ fermion bubbles and are shown diagrammatically in Fig.~\ref{fig:vertex}.

The term corresponding to pure QED, 
$\Pi^{(n)}_E$, was computed in Ref.~\cite{PalanquesMestre:1983zy}, 
and the pure-Yukawa contributions, $S^{(n)}_K$, $\Sigma^{(n)}_K$
and $V^{(n)}_K$, in Ref.~\cite{Alanne:2018ene}.
Their contribution to the $\beta$-functions, Eqs~\eqref{eq:coupledE} and~\eqref{eq:coupledK}, is
\begin{align}
    \label{eq:isolationE}
    \beta_E(K=0)=&E^2\left[\frac{2}{3}+\frac{1}{4N_f}\int_0^{2/3 E}\pi_E(t) \mathrm{d}t\right]+\mathcal{O}(1/N_f^2),\\
    \label{eq:isolationK}
    \beta_K(E=0) =& K^2\left[ 
	1 + \frac{1}{N_f} \left(\frac{3}{2}
	+ \int_0^K \xi_K(t) \mathrm{d}t \right)\right]+\mathcal{O}(1/N_f^2),
\end{align}
where
\begin{align}
    \pi_E(t)&=\frac{\Gamma(4-t)(1-t)\left(1-\frac{t}{3}\right)\left(1+\frac{t}{2}\right)}{\Gamma\left(2-\frac{t}{2}\right)^2
	\Gamma\left(3-\frac{t}{2}\right)\Gamma\left(1+\frac{t}{2}\right)},\\
\xi_K(t) &
=-\frac{\Gamma(4-t)}{\Gamma(2-\frac{t}{2})\Gamma(3-\frac{t}{2})\pi t}\sin\left(\frac{\pi t}{2}\right)
    \end{align}

The impact of the mixed contributions,
namely $\Pi_K^{(n)}$, and $S_E^{(n)}$, $\Sigma_E^{(n)}$, $V_E^{(n)}$, 
is evaluated in the next section.

\section{Mixed contributions}
\label{sec:renC}

In this section we derive the mixed contributions to the renormalization
constants for the photon self-energy, 
the fermion self-energy, the Yukawa vertex, and the scalar self-energy, 
and eventually compute the coupled $\beta$-functions.

\subsection{The Yukawa contribution to the QED $\beta$-function}
The Yukawa contribution to the photon self-energy (depicted in the second row of Fig.~\ref{fig:photonSE}),
is obtained by substituting  Eq.~\eqref{eq:pi0} in Eq.~\eqref{eq:Z_3def}.
We get
\begin{equation}\label{eq:Z_3K}
Z_3(K) = 
- \frac{E}{N_f}
\text{div}\left\{\sum_{n=1}^\infty
(Z_K K)^{n} \Pi_K^{(n+1)}(p^2,\epsilon)
\right\}.
\end{equation}
Notice that the diagrams involving a horizontal bubble chain differ from the 
corresponding ones for the scalar self-energy in
Fig.~\ref{fig:scalarSE} just by an overall factor $(2-d)$
coming from the algebra of the $\gamma$-matrices.
Altogether, we find
\begin{equation}\label{eq:pik}
 \Pi_K^{(n)}(p^2,\epsilon) = (-1)^{n-1} \frac{3}{4(d-1) n \epsilon^{n-1}}
 \pi_K(p^2,\epsilon,n),
\end{equation}
where $\pi_K(p^2,\epsilon,n)$ can be expanded as
\begin{equation}
 \pi_K(p^2,\epsilon,n) = 
 \sum_{j=0}^\infty \pi_K^{(j)}(p^2,\epsilon) (n \epsilon)^j ,
\end{equation}
with $\pi_K^{(j)}(p^2,\epsilon)$ regular for $\epsilon \rightarrow 0$.
Recalling that $Z_K = \left(1 - \frac{1}{\epsilon} K\right)^{-1} + \mathcal{O}(1/N_f)$,
we can evaluate $Z_3(K)$ from Eq.~\eqref{eq:Z_3K}:
\begin{equation}
\begin{split}
Z_3(K) & = 
- \frac{E}{N_f}
\text{div}\left\{\sum_{n=1}^\infty K^n
\sum_{i=0}^{n-1}
\left( \begin{array}{c} n  -1 \\ i \end{array} \right)
\frac{1}{\epsilon^i} \, \Pi_K^{(n-i+1)}(p^2,\epsilon) 
\right\} \\
& = - \frac{3 E}{4 N_f}
\text{div}\left\{\sum_{n=1}^\infty \frac{(-K)^n}{(d-1)\epsilon^{n}}
\sum_{j=0}^{n-1}\pi_K^{(j)}(p^2,\epsilon) \epsilon^j \right.\\
&\qquad\qquad\qquad\left.\times\sum_{i=0}^{n-1}
\left( \begin{array}{c} n  -1 \\ i \end{array} \right)
(-1)^i (n-i+1)^{j-1}
\right\} \\
& = - \frac{3 E}{4 N_f}
\text{div}\left\{\sum_{n=1}^\infty \frac{(-K)^n}{(d-1)\epsilon^{n}}
\pi_K^{(0)}(\epsilon) \frac{(-1)^{n+1}}{n(n+1)}
\right\} \\
& = - \frac{3 E}{4 N_f} \frac{1}{\epsilon}
\int_0^K \frac{ \pi_K^{(0)}(t)}{t-3} \left( 1 - \frac{t}{K}\right) \mathrm{d}t,
\end{split}
\end{equation}
where we used
\begin{equation}
\sum_{i=0}^{n-1}
\left( \begin{array}{c} n  -1 \\ i \end{array} \right)
(-1)^i (n-i+1)^{j-1} = \frac{(-1)^{n+1}}{n(n+1)} \delta_{j,0}, \quad j = 0,\dots,n-1
\end{equation}
and restricted ourselves to the $1/\epsilon$ pole.
The function $\pi_K^{(0)}$ is independent 
of $p^2$, as it should, and reads
\begin{equation}
\pi_K^{(0)}(t)=\frac{(t-2)(t-1)\Gamma(5-t)}{6\Gamma(3-\frac{t}{2})^2\pi t}\sin\left(\frac{\pi t}{2}\right).
\end{equation}

\begin{figure}
    \begin{center}
	\includegraphics[width=0.48\textwidth]{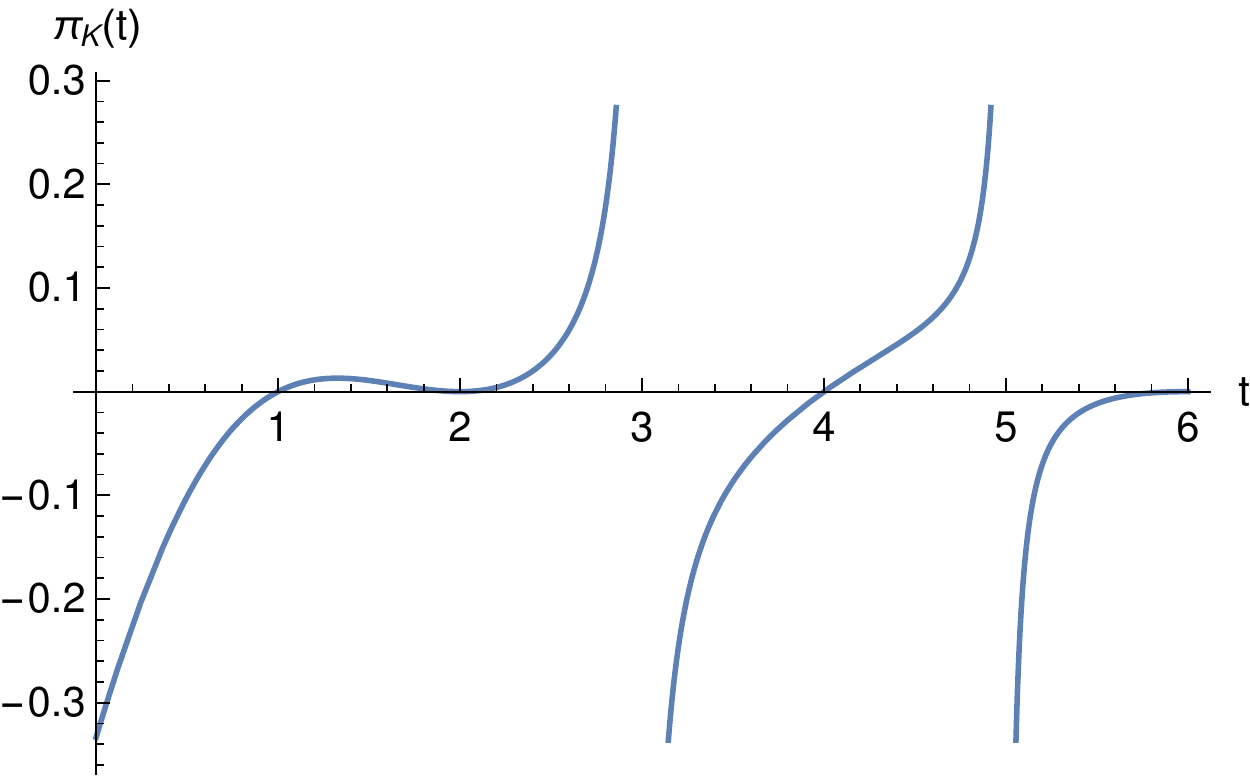}
    \end{center}
    \caption{The function $\pi_K(t)$ defined in Eq.~\eqref{eq:piK}.}
    \label{fig:piK}
\end{figure}

The contribution of $Z_3(K)$
to $\beta_E$,  Eq.~\eqref{eq:coupledE}, is found to be
\begin{equation}
\label{eq:betaEK}
\beta_E(K\neq0) = E^2 \frac{3}{4 N_f}
\int_0^K \pi_K(t) \mathrm{d}t.
\end{equation}
where we have defined
\begin{equation}
\label{eq:piK}
 \pi_K(t) \equiv \frac{ \pi_K^{(0)}(t)}{t-3}.
\end{equation}
We show the function $\pi_K(t)$ in Fig.~\ref{fig:piK}.
Since $\pi_K(t)$ has a first order
pole at $t=3$, the first singularity of
$\beta_E(K\neq0)$ occurs at $K=3$ and is a logarithmic one. 
The next singularity of $\pi_K(t)$ is found at $t=5$ (first order)
and would result in a logarithmic singularity
of $\beta_E(K\neq0)$ at $K=5$.

\subsection{The QED contribution to the Yukawa $\beta$-function}

The QED contribution to the fermion self-energy
and to the Yukawa vertex is closely related to the 
pure-Yukawa case. This is because
the gauge chain is equivalent to the Yukawa chain
besides an overall factor. 
In fact, $\Sigma_E^{(n)}$ and $V_E^{(n)}$ are related to
 $\Sigma_K^{(n)}$ and $V_K^{(n)}$
as
\begin{align}\label{eq:analog1}
 \Sigma^{(n)}_E(\slashed{p}) =& (d-2)
 \left(\frac{d-2}{d-1}\right)^{n-1} \Sigma_K^{(n)}(\slashed{p}),\\
\label{eq:analog2}
V^{(n)}_E (p^2) =&
  -d \left(\frac{d-2}{d-1}\right)^{n-1} {V_K^{(n)}}(p^2). 
\end{align}
The factors $(d-2)$ and $-d$ come from the algebra of the 
$\gamma$-matrices, while $\left(\frac{d-2}{d-1}\right)^{n-1}$
takes into account the difference in replacing
$\Pi_E^{(1)}$ with $S_K^{(1)}$.
Notice that $g_{\mu \nu}$ is the only relevant
Lorentz structure in the photon propagator,
since the $k_\mu k_\nu$ term do not contribute
to the $\beta$-function.

Making use the relations Eqs~\eqref{eq:analog1}
and~\eqref{eq:analog2},
$\Sigma_E^{(n)}$ and $V_E^{(n)}$ are expanded as
\begin{equation}
\Sigma^{(n)}_E(\slashed{p}) 
= (-1)^{n-1} \left( \frac{2}{3} \right)^n \frac{3}{ 4 n \epsilon^n} \sigma_E(p^2,\epsilon,n),
\end{equation}
\begin{equation}
 \sigma_E(p^2,\epsilon,n) = 
 \sum_{j=0}^\infty \sigma_E^{(j)}(p^2,\epsilon) (n \epsilon)^j,
\end{equation}
and
\begin{equation}
V^{(n)}_E (p^2) =  (-1)^{n-1} \frac{3}{n \epsilon^n} 
\left(\frac{2}{3}\right)^n v_E(p^2,\epsilon,n),
\end{equation}
\begin{equation}
 v_E(p^2,\epsilon,n) = 
 \sum_{j=0}^\infty v_E^{(j)}(p^2,\epsilon) (n \epsilon)^j.
\end{equation}
Using the one-loop result
$Z_E = \left( 1 - \frac{2}{3} E \right)^{-1} + \mathcal{O}(1/N_f)$,
and applying the same summation procedure as in
Ref.~\cite{Alanne:2018ene} for the fermion self-energy and the vertex, 
Eqs~\eqref{eq:Z_fdef} and~\eqref{eq:Z_Vdef} yield
\begin{equation}\label{eq:Z_fefinal}
Z_f(E) = - \frac{1}{N_f} \sum_{n=1}^\infty 
\text{div} \left\{(Z_E E)^n \Sigma_E^{(n)}(p^2,\epsilon) \right\}
= - \frac{1}{N_f} \frac{3}{4 \epsilon}
\int_0^{\frac{2}{3} E} \sigma_E^{(0)}(t) \mathrm{d}t,
\end{equation}
\begin{equation}\label{eq:Z_Vefinal}
Z_V(E) = - \frac{1}{N_f} \sum_{n=1}^\infty 
\text{div} \left\{(Z_E E)^n V_E^{(n)}(p^2,\epsilon) \right\} 
= - \frac{1}{N_f} \frac{3}{\epsilon}
\int_0^{\frac{2}{3} E} v_E^{(0)}(t) \mathrm{d}t,
\end{equation}
where we kept only the $1/\epsilon$ pole.
The functions $\sigma_E^{(0)}$ and $v_E^{(0)}$ are independent
of $p^2$, and are given by 
\begin{align}
 \sigma_E^{(0)}(t) =&  \frac{ 2 \Gamma(4-t) }{3 \pi \Gamma \left(1-\frac{t}{2}\right) 
 \Gamma \left(3-\frac{t}{2}\right) t}\sin\left( \frac{\pi t}{2}\right),\\
 v_E^{(0)}(t) =& \left(\frac{1 - \frac{t}{4}}{1- \frac{t}{2}}\right)^2 
 \sigma_E^{(0)}(t).
\end{align}

The QED contribution to the scalar self-energy
is shown in the second row of Fig.~\ref{fig:scalarSE}.
The diagrams involving a horizontal
gauge chain are related to the ones in the
pure-Yukawa case analogoursly to 
Eq.~\eqref{eq:analog1}.
Altogether, we find 
\begin{equation}
S_E^{(n)}(p^2,\epsilon) = (-1)^{n}
\left( \frac{2}{3} \right)^{n} 
\frac{27}{4 n(n-1)\epsilon^{n}} s_E(p^2,\epsilon,n),
\end{equation}
\begin{equation}
 s_E(p^2,\epsilon,n) = 
 \sum_{j=0}^\infty s_E^{(j)}(p^2,\epsilon) (n \epsilon)^j.
\end{equation}
The QED contribution in Eq.~\eqref{eq:Z_Sdef} is singled out as follows:
\begin{equation}\label{eq:Z_Se}
Z_S(E)  =  - K \text{div}\left\{ Z_f(E)^{-2}Z_V(E)^2 S_K^{(1)}(p^2,\epsilon)
+ \frac{1}{N_f} \sum_{n=1}^\infty (Z_E E)^n S_E^{(n+1)}(p^2,\epsilon) \right\}.
\end{equation}
To evaluate the right-hand side of Eq.~\eqref{eq:Z_Se}, we
closely follow the procedure in Ref.~\cite{Alanne:2018ene} for
the scalar self-energy:
\begin{align}
    Z_S(E)& = - \frac{K}{N_f}  \sum_{n=1}^\infty E^n \text{div}
	\left\{ \left( 1 - \frac{2}{3} \frac{E}{\epsilon} \right)^{-n}
	\left[ 2 S^{(1)}_F \left(
	\Sigma^{(n)}_E - V^{(n)}_E \right)
	+S_E^{(n+1)} \right] \right\} \nonumber\\
    \begin{split}
	& =- \frac{K}{N_f} \sum_{n=1}^\infty E^n \text{div}
	    \left\{ 
	    \sum_{i=0}^{n-1}
	    \left( \begin{array}{c} n - 1 \\ i \end{array} \right)
	    \left(\frac{2}{3}\right)^i
	    \frac{1}{\epsilon^i}\right.\\
	&\left.\hspace{3.3cm}\times\left[ 2 S^{(1)}_F \left(
	    \Sigma^{(n-i)}_E - V^{(n-i)}_E \right)
	    +S_E^{(n-i+1)} \right] \vphantom{\sum_i^n}\right\} 
    \end{split}\\
    & = - 3 \frac{K}{N_f} \sum_{n=1}^\infty 
	\left( - \frac{2}{3} E \right)^n
	\text{div} \left\{
	\frac{1}{\epsilon^n}
	\sum_{i=0}^{n-1}
	\left( \begin{array}{c} n - 1 \\ i \end{array} \right)
	(-1)^i \frac{\xi_E(p^2,\epsilon,n-i)}{(n-i)(n-i+1) \epsilon^{n+1}} 
	\right\},\nonumber
\end{align}
where we defined
\begin{equation}
\xi_E(p^2,\epsilon,n) = \epsilon (n+1) \,2\, S_F^{(1)}\left(v_E(p^2,\epsilon,n)
- \frac{1}{4} \sigma_E(p^2,\epsilon,n)\right) - \frac{3}{2}
s_E(p^2,\epsilon,n+1),
\end{equation}
and $S_F^{(1)}$ is the finite part of the one-loop bubble 
$S_K^{(1)}$.
\begin{figure}[t]
    \begin{center}
	\includegraphics[width=0.48\textwidth]{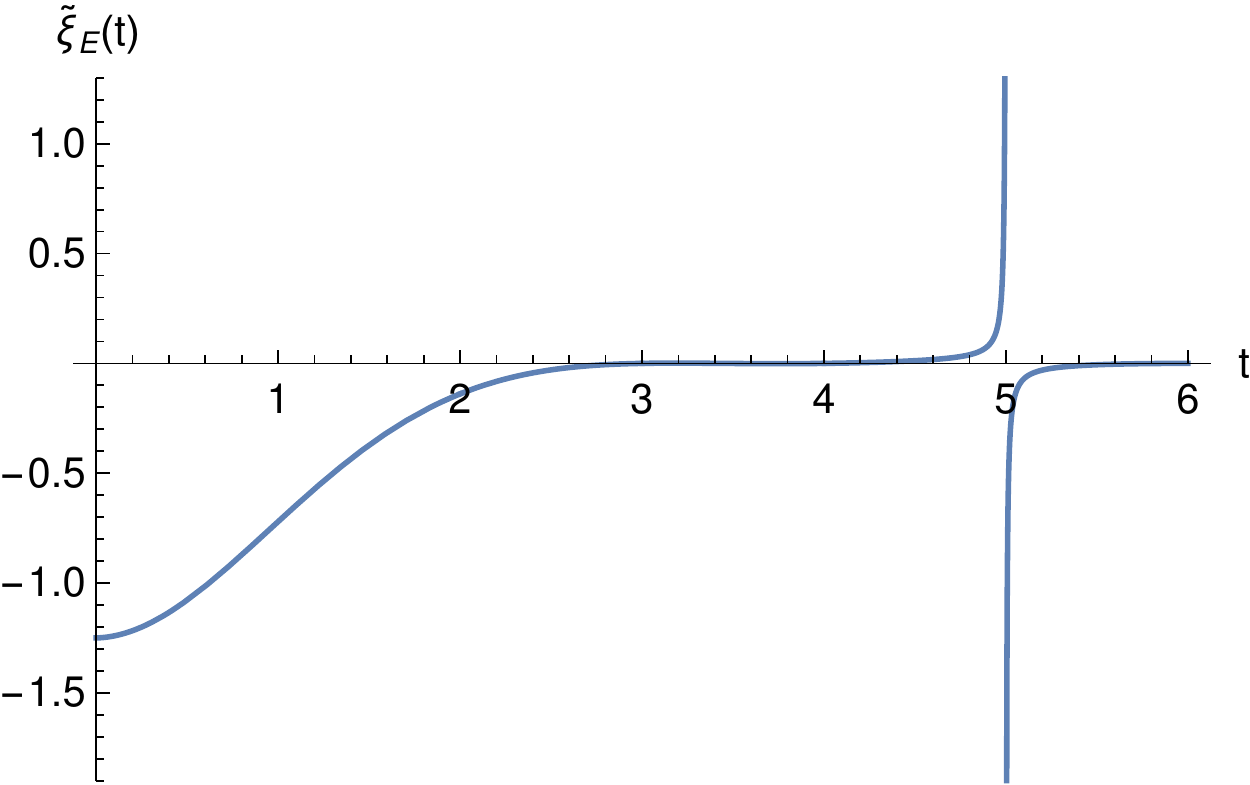}
	\quad	\includegraphics[width=0.48\textwidth]{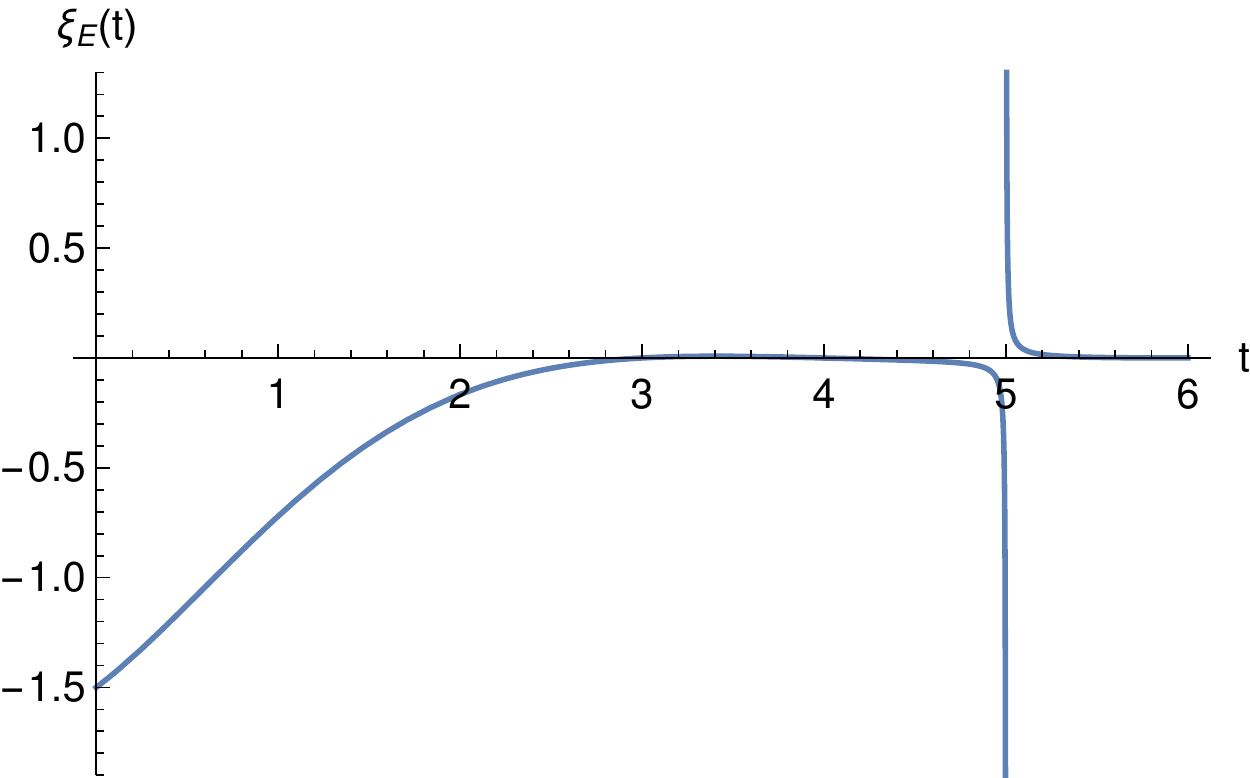}
    \end{center}
    \caption{The functions $\tilde{\xi}_E(t)$ (left panel) and $\xi_E(t)$ (right panel) 
	defined in Eqs~\eqref{eq:xiEtilde}  and~\eqref{eq:xiE}, respectively.}
    \label{fig:xis}
\end{figure}
Then, by expanding
\begin{equation}
\xi_E(p^2,\epsilon,n-i) = \sum_{j=0}^\infty
\epsilon^j (n-i+1)^j \xi_E^{(j)}(p^2,\epsilon),
\end{equation}
and using
\begin{equation}
\sum_{i=0}^{n-1} 
\left( \begin{array}{c} n -1 \\ i \end{array} \right) 
(-1)^i \frac{(n-i+1)^{j-1}}{n-i}
= \begin{cases}
\frac{(-1)^{n+1}}{n+1} & j=0 \\ 
\frac{(-1)^{n+1}}{n} & j=1,..,n
\end{cases},
\end{equation}
we can further simplify the expression to
\begin{equation}\begin{split}
Z_S(E) = & 3 \frac{K}{N_f} \sum_{n=1}^\infty
\left( \frac{2}{3} E \right)^n
\text{div} \left\{
\frac{1}{(n+1)\epsilon^{n+1}}
\xi^{(0)}_E(p^2,\epsilon)
+ \frac{1}{n \epsilon^{n+1}}
\sum_{j=1}^\infty  \xi_E^{(j)}(p^2,\epsilon)\epsilon^j
\right\} \\
=& 
 \frac{9 K}{2 E N_f} \sum_{n=2}^\infty
\left( \frac{2}{3} E \right)^{n}
\text{div} \left\{
\frac{1}{\epsilon^{n}}
\left(
\frac{\xi_E^{(0)}(p^2,\epsilon)}{n}
+
\frac{\xi_E(p^2,\epsilon,0) - \xi^{(0)}_E(p^2,\epsilon)}{n-1}\right)
\right\}\\
= & \frac{ 9 K}{2 E N_f}
\frac{1}{\epsilon}
\int_0^{\frac{2}{3}E} \left( \xi_E^{(0)}(t) - \xi_E^{(0)}(0)
+ \frac{2}{3} \frac{\xi_E(p^2,t,0)- \xi_E^{(0)}(t)}{t}  E \right) \mathrm{d}t, 
\end{split}
\end{equation}
where we kept the $1/\epsilon$ pole only. The function $\xi_E(p^2,t,0) = \lim_{n\to0} \xi_E(p^2,t,n)$ has to be independent
of $p^2$ for the consistency of the computation. This is indeed the case: we checked
that 
\begin{equation}
\frac{3}{2} s_E(p^2,t,1) =
2\left(1 + t \, S^{(1)}_F(p^2,t)\right) 
\left( v_E^{(0)}(t) - \frac{1}{4}\sigma_E^{(0)}(t) \right),
\end{equation}
and therefore
\begin{equation}
\xi_E(p^2,t,0)= - 2 v_E^{(0)}(t) + \frac{1}{2} \sigma_E^{(0)}(t) 
\equiv \xi_E(t).
\label{eq:xiE}
\end{equation}
Finally, 
we find:
\begin{equation}\label{eq:Z_Sefinal}
Z_S(E) 
= \frac{3 K}{\epsilon \, N_f} \left\{ \frac{3}{2 E}
\int_0^{\frac{2}{3}E} \left( \xi_E^{(0)}(t) - \xi_E^{(0)}(0) \right) \mathrm{d}t
+ 
\int_0^{\frac{2}{3}E} \frac{\xi_E(t) - \xi_E^{(0)}(t)}{t} \mathrm{d}t
\right\}.
\end{equation}
With Eqs~\eqref{eq:Z_fefinal}, \eqref{eq:Z_Vefinal} and
\eqref{eq:Z_Sefinal} at hand, we can compute the QED contribution
to the Yukawa $\beta$-function:
\begin{equation}\label{eq:betaKE}
\beta_K ( E \neq 0 ) =  - \frac{3 K^2}{N_f}
\left\{
\int_0^{\frac{2}{3}E } \tilde{\xi}_E(t) \mathrm{d}t
+ \frac{3}{2}
+ \left(1- \frac{2 E}{3 K} \right) \xi_E\left(\frac{2}{3} E\right) \right\}.
\end{equation}
where we have defined 
\begin{equation}
 \tilde{\xi}_E(t) \equiv \frac{\xi_E(t)-\xi_E^{(0)}(t)}{t}. 
 \label{eq:xiEtilde}
\end{equation}
The functions $\xi_E(t)$ and $\tilde{\xi}_E(t)$
are explicitly given by
\begin{align}
 \xi_E(t) & = - \frac{ 2 (t-3)^2 \Gamma(2-t)}{3 \Gamma\left(2-\frac{t}{2}\right) 
 \Gamma\left(3-\frac{t}{2}\right)\pi t}\, \sin\left(\frac{\pi t}{2}\right), \\
 \tilde{\xi}_E(t)& =\frac{(15+t-5t^2+t^3)\Gamma(4-t)}
 {9(t-2)\Gamma\left(2-\frac{t}{2}\right)\Gamma(3-\frac{t}{2})\pi t}
 \sin\left( \frac{\pi t}{2}\right).
\end{align}

We plot the functions $\tilde{\xi}_E(t)$ and $\xi_E(t)$ in Fig.~\ref{fig:xis}. 
The first singularity of $\beta_K(E\neq0)$ is at $E=15/2$
and consists of a first-order pole coming from
$\xi_E(t)$ 
plus a logarithmic 
singularity arising from the integration of
$\tilde{\xi}_E(t)$, both at $t=5$.

\section{The coupled system}
\label{sec:beta}

Here we summarize and discuss our results for the coupled system.
Combining Eqs~\eqref{eq:isolationE} and~\eqref{eq:isolationK} with the new results
in Eqs~\eqref{eq:betaEK} and \eqref{eq:betaKE}, we obtain 
\begin{align}\label{eq:coupledfinal}
 \frac{\beta_K}{K^2} = &
1 - \frac{3}{N_f} \left\{1 -
\frac{1}{3}\int_0^K \xi_K(t) \mathrm{d}t
+ \int_0^{\frac{2}{3}E} \tilde{\xi}_E(t) \mathrm{d}t
+ \left(1 -\frac{2 E}{3 K} \right) \xi_E\left(\frac{2}{3}E\right) \right\},
\\ 
\frac{\beta_E}{E^2} = & \frac{2}{3} + \frac{1}{4 N_f} \left\{ 
\int_0^{\frac{2}{3} E} \pi_E(t) \mathrm{d}t + 
3 \int_0^K \pi_K(t) \mathrm{d}t \right\}.
\end{align}
Near the Gaussian fixed point, these can be expanded as
\begin{align}
    \label{eq:bEexp}
    \begin{split}
    \beta_E=&\frac{2}{3}E^2+\frac{1}{2N_f}E^3-\frac{1}{4N_f}E^2K-\frac{11}{72N_f}E^4
	+\frac{7}{32N_f}E^2K^2\\
	&-\frac{77}{1944N_f}E^5-\frac{3}{64 N_f}E^2K^3	+\dots
	\end{split}
	\\
	\begin{split}
    \beta_K=&\left(1+\frac{3}{2N_f}\right)K^2-\frac{3}{N_f}EK-\frac{3}{2N_f}K^3+\frac{5}{4N_f}E K^2+\frac{5}{6N_f}E^2K\\
	&+\frac{7}{16N_f}K^4-\frac{1}{2N_f}E^2K^2+\frac{35}{108N_f}E^3K\\
	&+\frac{11}{96 N_f}K^5+\frac{1}{3888N_f}\left(-1625+1296\zeta_3\right)E^3 K^2 +\frac{1}{648N_f}(83-144\zeta_3)E^4K\dots
    \end{split}
    \label{eq:bKexp}
\end{align}
We have checked that the expansions agree with the known four-loop 
results~\cite{Machacek:1983tz,Machacek:1983fi,Pickering:2001aq,Chetyrkin:2012rz,Zerf:2018csr}
in the leading order in $N_f$. Furthermore, the $-\frac{2 E}{3 K} \xi_E\left(\frac{2}{3}E\right)$ part in
the last term of Eq.~\eqref{eq:coupledfinal} corresponds to the result of Refs~\cite{Kowalska:2017pkt,Antipin:2018zdg}, and we have 
checked that our result agrees with those.

The first singularity of the pure-QED $\beta$-function is located at $E=15/2$, 
whereas for the pure-Yukawa case it occurs at $K=5$. 
These known singularities are now accompanied by the ones 
from the mixed contributions, Eqs~\eqref{eq:betaEK}
and \eqref{eq:betaKE}. As we noticed in 
Section~\ref{sec:renC}, $\beta_E(K\neq0)$
has the first singularity at $K=3$,
while $\beta_K(E\neq0)$ at $E=15/2$.
The former, similarly to the pure gauge and Yukawa cases, 
is a logarithmic singularity, whereas the latter is
a pole of first order.

The $\mathcal{O}(1/N_f)$ coupled system 
has only the three already known fixed points: 
the Gaussian fixed point, and the pure-QED (near $E=15/2$) 
and pure-Yukawa (near $K=3$) fixed points.

\begin{figure}
    \begin{center}
	\includegraphics[width=0.45\textwidth]{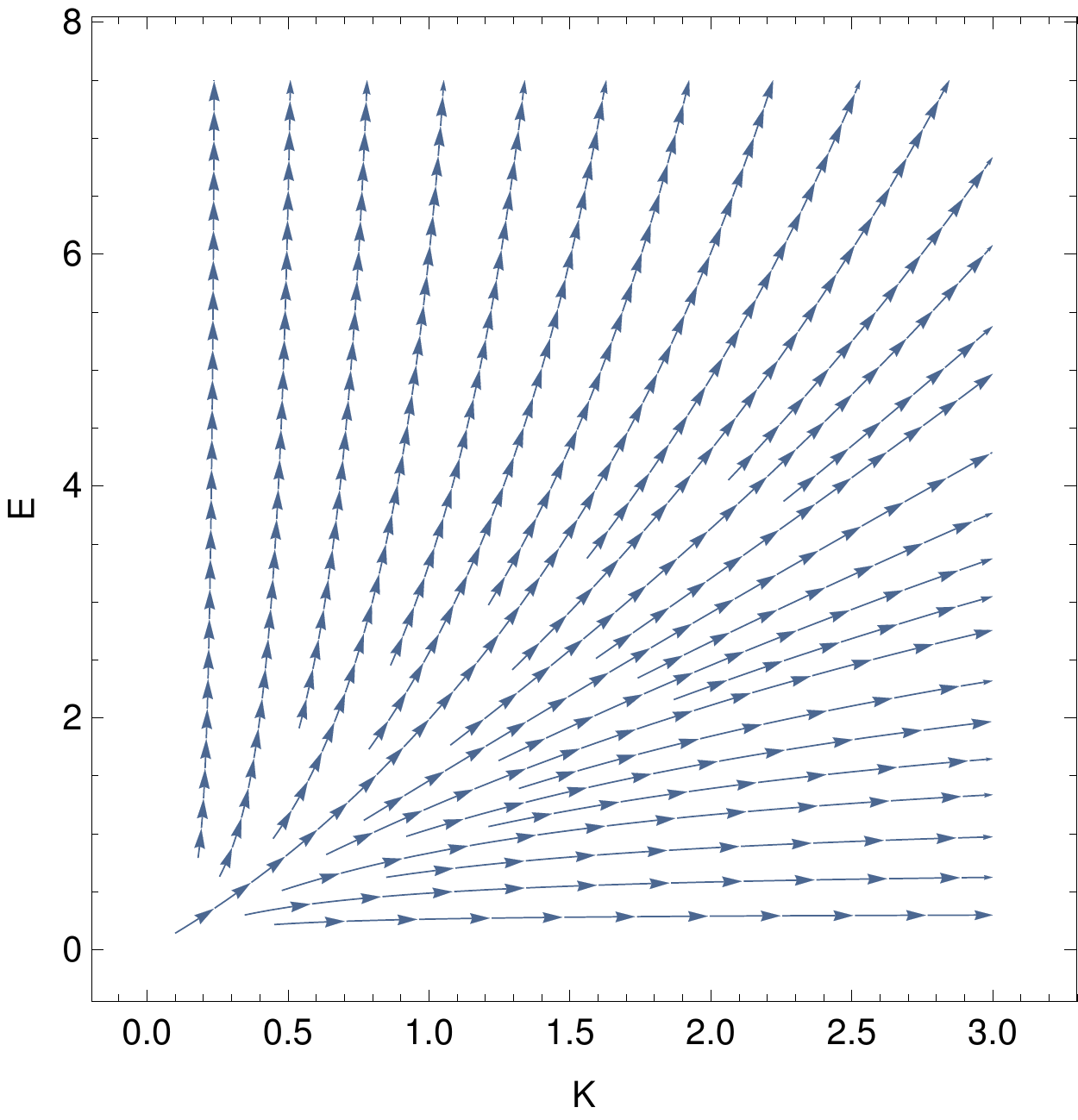}
    \end{center}
    \caption{The flow diagram for the coupled system with $N_f=30$. 
    The arrows point towards UV.}
    \label{fig:flow30}
\end{figure}

We show the flow diagram for $N_f=30$ outside the vicinity of the singularities
in Fig.~\ref{fig:flow30}.
Near $K=3$, the logarithmic singularity in $\beta_E$ arising from
$\pi_K(t)$ dominates making the gauge coupling to increase and approach the value $E=15/2$. 
Near $E=15/2$, however, $\beta_K$ has a pole arising from $\xi_E(t)$
eventually dominating the flow, and driving the Yukawa coupling to zero near $E=15/2$.
The flow  may be extended setting $K\equiv0$ and switching to pure-QED, 
so that the gauge coupling reaches the fixed point as $E\to 15/2$ in the UV.

\section{Conclusions}
\label{sec:concl}
We have computed the leading $1/N_f$ mixed contributions 
for the  $\beta$-functions for abelian gauge-Yukawa theory with $N_f$ fermion flavours
coupling to a gauge-singlet real scalar. Together with the known results for the 
pure-QED and pure-Yukawa cases, this allows the study of the abelian gauge-Yukawa system.

The flow in the interacting theory leads to the vanishing Yukawa coupling near the gauge coupling value $E=15/2$
due to the peculiar interplay of the singularities.
However, the gauge $\beta$-function is still positive around $(K,E)=(0,15/2)$, and $E$ keeps growing before eventually reaching 
the fixed point due to the known a logarithmic singluarity near $E=15/2$. 

Our work extends the previous results towards a more complete picture of gauge-Yukawa theories in
the large-$N_f$ limit.

\section*{Acknowledgments}
We thank John Gracey for valuable comments.


\bibliography{refs.bib}
\bibliographystyle{JHEP}

\end{document}